\documentclass[preprint]{aastex}
\newcommand{\boldmathsymbol}[1]{\mbox{\boldmath $#1$}}
\newcommand{\vect}[1]{\boldmathsymbol{#1}}
\newcommand{\eq}[1]{equation (\ref{#1})}
\newcommand{\vhat}[1]{\boldmathsymbol{\hat{#1}}}
\newcommand{\p}{\vhat{p}}
\newcommand{\q}{\vhat{q}}

\newcommand{\y}{\lambda}
\newcommand{\cross}{\boldmathsymbol{\times}}
\renewcommand{\d}{\mbox{d}}
\newcommand{\D}{\mbox{\bf d}}
\newcommand{\Dy}{\mbox{d}\y}
\newcommand{\R}{{\cal R}}
\newcommand{\F}{{\cal F}}
\newcommand{\teff}{\tau_{\mbox{\tiny eff}}}
\newcommand{\bra}{\langle}
\newcommand{\ket}{\rangle}

\begin{document}

%{{{ Title page and ABSTRACT 

\thispagestyle{empty}

\title{The Formal Underpinnings of the Response Functions used in X-Ray
Spectral Analysis}
\author{John E. Davis}
\email{davis@space.mit.edu}
\affil{Center for Space Research, Massachusetts Institute of Technology,
       Cambridge MA, 02139}

\begin{abstract}
  This work provides an in-depth mathematical description of the
  response functions that are used for spatial and spectral analysis
  of X-ray data.  The use of such functions is well-known to anyone
  familiar with the analysis of X-ray data where they may be
  identified with the quantities contained in the Ancillary
  Response File (ARF), the Redistribution Matrix File (RMF), and the
  Exposure Map.  Starting from first-principles, explicit mathematical
  expressions for these functions, for both imaging and dispersive
  modes, are arrived at in terms of the underlying instrumental
  characteristics of the telescope including the effects of pointing
  motion.  The response functions are presented in the context of
  integral equations relating the expected detector count rate to the
  source spectrum incident upon the telescope.  Their application to
  the analysis of several source distributions is considered.  These
  include multiple, possibly overlapping, and spectrally distinct
  point sources, as well as extended sources.  Assumptions and
  limitations behind the usage of these functions, as well as their
  practical computation are addressed.
\end{abstract}

\keywords{methods: data analysis, analytical --- X-rays: spectra}

%\maketitle

%}}}

\section{Introduction} %{{{

It is a basic tenet of X-ray spectral analysis that the
source flux incident at the telescope is related to the observed count rate
through an integral equation involving the effective area of the
telescope.  The most commonly accepted technique for dealing with this
equation involves the use of spectral analysis programs such as
\verb|xspec| \citep{xspecmanual}.  The effective area is input into
these programs via a file called the Ancillary Response File, or ARF.
In addition, the energy resolution of the detector is specified by the
Redistribution Matrix File, or RMF. 
%To complement the rigid,
%well-specified format specifications for these files \citep{ogiparf}
This work presents formal descriptions of the quantities
embodied by the ARF and RMF in terms of the underlying instrumental
responses, making a clear connection between the incident source
flux and the observed count rate.

Roughly speaking, the effective area of an X-ray telescope composed of
a mirror and a detector is more or less the product of the effective
area of the mirror with the quantum efficiency (QE) of the detector.
However, the ARF, which relates observed counts to a source flux, also
depends upon observation-dependent quantities such as the detailed
aspect history of the telescope, its point spread function (PSF), and
upon details of the analysis itself, e.g., the filtering and binning
of the observed data.

With the advent of the Chandra X-ray Observatory \citep{chandra},
all of the subtleties introduced by the shape of the PSF and the
telescope aspect motion were deemed important for the computation of
an ARF. Chandra is well calibrated \citep{chandracal} and, with its
unprecedented combined spectral and spatial response, a precise
definition of the ARF that incorporates the effects of spacecraft
motion and the PSF is necessary in order to perform spectral analysis
at the highest resolution of the instrument.

The main goal of this work is to present an explicit first-principles
derivation of the ARF by including the proper treatment of telescope
motion (e.g., dither) and the PSF, as well as any filtering of the
data.  The bottom-up approach taken here necessarily implies that a
meaningful and consistent derivation can be achieved by considering
the role of the ARF in spectral analysis.  As a result, the ARF is
presented in the context of integral equations that connect the
incident X-ray flux to an expected count rate.
%Both dispersive and non-dispersive modes are considered.

Traditionally, the ARF and RMF have been used primarily for the
analysis of spectral image data.  An important aspect of this work is
to extend this approach to the analysis of dispersive spectral data,
such as data obtained by Chandra or Newton. To this end, definitions
of an ARF and an RMF are presented that are suitable for dispersive
data analysis, and which may be utilized by existing spectral analysis
software.

One of the original motivations for this work was the need to create a
related object, an exposure map, for use in the analysis of data
obtained by the Chandra X-ray Observatory.  This paper also gives a
rigorous definition of the exposure map and discusses some of its uses
and its limitations in spectral image analysis.  The resulting
definition is consistent with current use, intuition, and physics.

The next section contains a discussion of the general response of an
X-ray telescope and also serves to introduce the notation and
conventions used throughout this paper.  Although originally inspired
by the need to create ARFs and exposure maps for Chandra, the
presentation has been kept as general as possible without focusing on
any particular telescope or instrument. A derivation of the imaging
ARF follows in section \ref{sec:arfdef} where its application to
several problems is considered.  These include the problem of multiple
overlapping point sources.  Section \ref{sec:exposuremap} contains a
definition of the exposure map and explores its use as well as its
limitations in dealing with extended sources.  The definition of a
dispersive ARF and RMF that are suitable for use in the analysis of
dispersive spectral data are given in section \ref{sec:garf}.
Following the summary of the paper is an appendix that considers the
practical computation of these objects.

%}}}

\section{Telescope Response} %{{{
\label{sec:telescoperesponse}

Let $S(\y,\p)d{\y}\;\D\p$ represent the number of photons per unit
area per unit time incident upon the telescope with directions that
lie in the cone of directions between $\p$ and $\p+\D\p$, and whose
wavelengths lie between $\y$ and $\y+d{\y}$.  This source flux is
assumed to be time-independent\footnote{%
    If the source flux varies in time such that its time-dependence is
    uncorrelated with the spatial and spectral shape, as is often the
    case, then one can always factor out the time-dependence and
    handle it separately.  This technique is discussed in more detail
    below. The treatment of more complex time-dependent sources that
    do not admit this factorization is beyond the scope of this
    paper.
}. %
Similarly, let $S_D(h,\sigma,t)\d{\sigma}$ represent the {\em
expected} number of counts per unit time, in pulse-height bin $h$,
within a region of area $\d{\sigma}$ at the position $\sigma$ on the
detector.  (Although in this paper, $h$ is a discrete quantity that
represents a pulse-height channel, one could easily generalize $h$ to
a continuous quantity such as a voltage by introducing the
infinitesimal $\d{h}$ and replacing $\sum_h$ by $\int\d{h}$.)

In general, the count rate will be time-dependent even if the source
flux does not vary with time. This time-dependence arises from several
effects, including but not limited to telescope pointing motion (e.g.,
pointing wobble or dither), detector electronics, telemetry
saturation, and thermal expansion effects that cause the movement of
individual telescope subsystems with respect to one another.

As used here, pointing is the measured, generally periodic movement of
a coordinate system attached to the optical axis with respect to a
coordinate system fixed with respect to the sky.  These two coordinate
systems are related by a time-dependent rotation matrix $\R(t)$ which
completely characterizes the pointing.  It is assumed that complete
knowledge of $\R(t)$ is available from, e.g., an on-board aspect
system, at some level of accuracy (see section
\ref{sec:telescoperesponse}).

%For instance
%ASCA was not purposely dithered; nevertheless data from its
%star-tracker system may be used to determine its instantaneous
%pointing.

%The time dependent effects caused by the detector electronics and
%telemetry saturation are usually incorporated into what are called
%``Good-Time Intervals''.

In the coordinate system fixed with respect to optical axis, the
source flux will appear to be time-dependent according to
\begin{equation} 
   S(\y, \p, t) = S(\y, \R^{-1}(t)\cdot\p).
\end{equation}
This equation simply expresses the fact that an observer fixed to the
telescope will see a time dependent source and that this induced time
dependence is a direct consequence of telescope motion.
In the telescope coordinate system, the source flux and the count rate are
related to one another via the equation
\begin{equation}
   \label{rtdef}
   S_D(h, \sigma, t) = \int \Dy \int \D\p \; R_T(h, \y, \sigma, \p, t)
        S(\y, \R^{-1}(t)\cdot\p),
\end{equation}
which defines the total {\em response} $R_T(h,\y,\sigma,\p,t)$ of the
telescope. It is an extremely complicated function that incorporates
all elements of the telescope such as the detector and its
electronics, the mirror, and diffraction gratings, if present. Since
the units of $S_D(h,\sigma,t)$ are counts per unit detector area per
unit time, and the units of $S(\y,\p)\d{\y}$ are photons per unit
aperture area per unit time, it follows that the response
$R_T(h,\y,\sigma,\p,t)$ is a unitless quantity (counts/photon $\times$
aperture area/detector area).

The mathematical form of \eq{rtdef} represents a linear mapping from
$S$ to $S_D$.  Hence, strictly speaking, \eq{rtdef} is valid only when
one can neglect non-linear effects such as local gain depression or
pile-up, i.e., a non-linear effect caused by the finite time
resolution of the detector.  Pile-up can be treated, at least in
principle, by making the response $R_T(h,\y,\sigma,\p,t)$ a function
of the incident flux $S(\y,\p)$.  However, this topic is beyond the
scope of the present work and will be addressed
elsewhere \citep{davispileup}.

The explicit time-dependence of $R_T(h,\y,\sigma,\p,t)$ is due to
effects associated with telemetry saturation, thermal expansion, and
so on.  It is assumed that all but the time-dependence associated with
internal movement of the telescope subsystems may be encapsulated in a
function $T(\sigma,t)$ that factors out of $R_T(h,\y,\sigma,\p,t)$,
i.e.,
\begin{equation}
    R_T(h, \sigma, \y,\p, t) =
      T(\sigma, t) R(h, \sigma, \y, \p, t).
\end{equation}
In this equation the residual time-dependence of the response function
$R(h,\sigma,\y,\p,t)$ depends only upon the relative movements of the
individual telescope subsystems.  The function $T(\sigma, t)$ can be
thought of as representing the the so-called ``good-time intervals'',
or GTIs.  However, $T(\sigma,t)$ could play a more general role than
this because it could include the effects of bad-pixels\footnote{%
    As described below, the effects of static bad-pixels are assumed to be
    contained in the detector response function $D(h,\sigma,\y)$.
}, %
which themselves may be time-dependent, as well as any dead-time
factors associated with telemetry saturation. It should also be noted
that although the incident source flux has been assumed to be
independent of time, $T(\sigma,t)$ could encompass time-dependent
variations in the source flux as long as the spectral shape itself
does not depend upon time.  In this case, only the amplitude of the
flux is time-dependent and this time-dependence may be factored out of
the incident source flux and into the function $T(\sigma,t)$.  Further
exploration of this possibility is left to the reader.  Suffice it to
say that the actual values that $T(\sigma,t)$ take on are not
important in what follows.  Hence,
\eq{rtdef} can be written
\begin{eqnarray} \label{response}
   S_D(h, \sigma, t) &=& T(\sigma, t) \int \Dy \int \D\p \;
      R(h, \sigma, \y, \p, t) S(\y, \R^{-1}(t)\cdot\p) \\
      &=& T(\sigma, t) \int \Dy \int \D\p \;
      R(h, \sigma, \y, \R(t)\cdot\p, t) S(\y, \p) ,  \nonumber
\end{eqnarray}
where the last form follows from the change of variables
$\p\longrightarrow\R(t)\cdot\p$ and noting that because the matrix
$\R(t)$ is orthogonal, the Jacobian of the transformation is unity.

Assuming that the response depends upon several {\em independent}
telescope subsystems, it may be factored it into subsystem-dependent
pieces.  The specific form of the factorization will depend upon the
actual physical relationships between the various subsystems.  For a
prototypical X-ray telescope with a focusing mirror at the aperture of
the telescope that focuses X-rays onto a position-sensitive detector,
an appropriate factorization is given by
\begin{equation} \label{imagingrsp}
  R(h, \sigma, \y, \p,t) =
     D(h, \sigma, \y) 
       \int\D\p' \; \delta(\sigma - \sigma(\p',t))
       \F(\y,\p',\p) M(\y,\p).
\end{equation}
A different factorization may have to be use to describe the response
of some other type of telescope (see section \ref{sec:garf} for the
factorization needed to to describe the presence of a diffraction
grating). In this equation, $M(\y,\p)$ is the off-axis effective area
of the mirror, and $D(h,\sigma,\y)$ represents the probability that a
photon with wavelength $\y$ at position $\sigma$ on the detector will
give rise to a pulse-height $h$.  The term involving the delta
function symbolizes the passage of a photon with direction $\p'$ from
the mirror to the position $\sigma$ on the detector via the coordinate
transformation $\sigma(\p',t)$.  In general, this function is
time-dependent and represents any relative motion that exists between
the mirror and the detector.

\begin{figure}
\epsscale{0.8}
\plotone{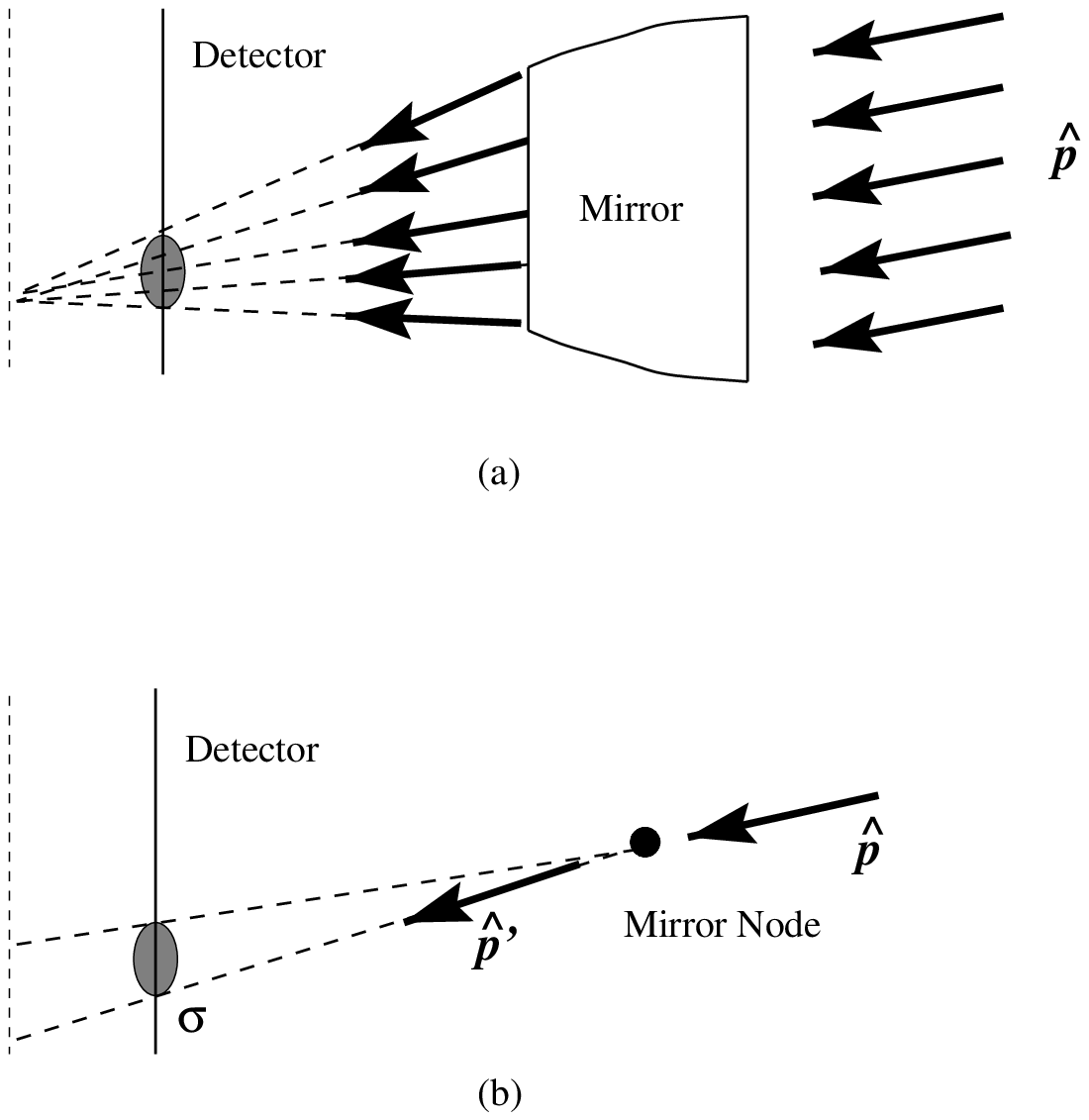}
\caption{%
\label{fig:psf}
Figure illustrating how the PSF function $\F(\y,\p',\p)$ may be used
to accurately model a realistic mirror of finite extent.  The top
portion of the figure shows a ``realistic'' mirror with the detector in a
de-focused position. In this position photons from a distant point
source enter the mirror at the front and exit the mirror at the back,
and are focused towards a point behind the detector causing the point
source to appear out of focus. The bottom part of the figure shows
that this effect can be modeled in terms of the concept of a ``mirror
node'', where rays enter and leave the same point with the appropriate
redistribution $\F(\y,\p',\p)$ of ``incoming'' photons with direction
$\p$ into ``outgoing'' ones with direction $\p'$. In particular, note
that the same PSF function appropriate for the de-focused position of
the detector cannot be used to model the image when the detector is at
the focal position; there a different PSF will be required.  In other
words, for $\F(\y,\p'.\p)$ to accurately model the mirror, it also
depends {\em implicitly} upon the location of the detector with
respect to the focal surface. }
\end{figure}

The PSF (Point Spread Function) of the telescope is represented by the
function $\F(\y,\p',\p)$, which is assumed to satisy the normalization
condition
\begin{equation}
   1 = \int \D\p' \F(\y,\p',\p).
\end{equation}
Its definition is based upon the idea (see figure \ref{fig:psf}) that the
mirror itself (e.g., one of a typical type-I Wolter design) may be
modeled by the appropriate probability distribution for photons to
enter and leave the mirror at a single point (the so-called ``mirror
node''), with an effective area given by $M(\y,\p)$.  As indicated in
figure \ref{fig:psf}, the PSF defined in this way will also depend
(implicitly) upon the position of the detector and its relationship to
the focal surface.  For simplicity, in the following it will be
assumed that the variations in the movement of the detector with
respect to the mirror during the course of an observation are small
enough that the PSF may be regarded as independent of time to
sufficient accuracy.  The reader should note that a {\em perfect} PSF
defined in this sense is represented by
\begin{equation}
   \F_{\mbox{\tiny perfect}} (\y,\p',\p) = \delta(\p'-\p).
\end{equation}

%The function $\F(\y,\p',\p)$ represents the PSF (Point Spread
%Function) of system and is the probability density for a photon to
%deviate from its idealized direction $\p$ to a direction $\p'$. It is
%important to appreciate the fact that the PSF introduced this way is
%not simply a property of the mirror even though the form of the
%factorization suggests that it is.  For one thing, a real mirror such
%as one of a typical Wolter type-I design is an extended object that
%must necessarily change the direction of an incident photon in a
%manner that depends upon where the photon intercepts the surface of
%the mirror.  The function $\F(\y,\p',\p)$ actually represents the
%deviation of the optical system from that of an idealized one modeled
%as an infinitesimally small thin lens where photons enter the mirror
%at a single point and exit the mirror at the same point, with an
%effective area given by $M(\y,\p)$ (see figure \ref{fig:psf}). Hence,
%the PSF will also depend upon the position of the detector and its
%relationship to the focal surface.  For these reasons, it is best to
%think of $\F(\y,\p',\p)$ as the total system PSF.

Generally speaking, the detector response $D(h,\sigma,\y)$ can be
factored into a QE (Quantum Efficiency) $Q(\sigma,\y)$ and a redistribution
function $D_R(h, \sigma,\y)$:
\begin{equation} 
  \label{rmfdef}
   D(h, \sigma, \y) = D_R(h, \sigma, \y) Q(\sigma, \y).
\end{equation}
The function $D_R(h, \sigma, \y)$ is known as the redistribution
matrix function, or RMF \citep{ogiparf}, and represents a mapping, or
redistribution, from wavelength to pulse-height by the detector.
Without any loss of generality, it is assumed to be normalized to
unity via
\begin{equation}
   1 = \sum_h D_R(h, \sigma, \y).
\end{equation}
Alternatively, the quantum efficiency may be defined by
\begin{equation}  \label{qe}
   Q(\sigma,\y) = \sum_h D(h, \sigma, \y).
\end{equation}
In general, as indicated here the RMF varies with position on the
detector, although many applications assume a spatially constant RMF.
In contrast, the quantum efficiency function $Q(\sigma,\y)$ is assumed
to contain the effects of (static) bad pixels, detector dead regions,
detector boundaries, and so on, all of which cause it to vary
spatially.

Using the response function given by \eq{imagingrsp}, the
expected count rate from the source $S(\y,\p)$ is seen to be
\begin{eqnarray}
\label{expectrate}
     S_D(h, \sigma, t) &=&
        T(\sigma,t) 
           \int \d\y \int\D\p \; 
              R(h, \sigma, \y, \p_t, t) S(\y,\p) \\
           &=& \nonumber
        T(\sigma,t) 
           \int \d\y\;
               D(h,\sigma,\y) \int\D\p'\;
              \delta(\sigma-\sigma(\p',t)) 
                 \int\D\p\; \F(\y,\p',\p_t) M(\y,\p_t) S(\y,\p),
\end{eqnarray}
where, for notational simplicity, $\p_t$ symbolizes $\R(t)\cdot\p$.

Telescope pointing motion will cause events to appear spatially mixed
together when expressed as an image in detector coordinates.  For this
reason, it is preferable to work in the sky coordinate system where
one can remove the effects of the motion by projecting the events to
the sky in the appropriate manner.  Thus, we define an {\em
aspect-corrected} count rate via
\begin{equation} \label{aspectdata}
  S_A(h,\p,t) 
    = J(\sigma(\p_t,t),\p_t,t) S_D(h, \sigma(\p_t,t), t),
\end{equation}
where $J(\sigma,\p,t)$ is the instantaneous Jacobian of the
transformation from the detector coordinate $\sigma$ to the sky
coordinate $\p$ via the inversion of $\sigma(\p_t,t)$.  Physically,
the Jacobian represents the stretching or magnification of an element
of area on the detector as it appears in the sky.

By making use of the well known change of variable formula
for delta functions expressed in the form
\begin{equation}
   \label{deltacov}
   \delta(\sigma(\p_t,t) - \sigma(\p', t))
     = \frac{\delta (\p' - \p_t)}{J(\sigma(\p_t,t),\p_t,t)} ,
\end{equation}
one can show that the aspect-corrected count rate is
\begin{equation}
   \label{aspdatrelat}
     S_A(h,\p',t) = T(\sigma(\p'_t,t),t) 
           \int \d\y\; 
              D(\sigma(\p'_t,t),h,\y) \int \D\p\; \F(\y,\p'_t,\p_t)
                M(\y,\p_t) S(\y,\p).
\end{equation}
The above equation assumes that complete knowledge of the aspect
history is available in order to perform the aspect correction.  In
general, there will be uncertainties in the aspect solution which in
turn leads to spatial uncertainties in the aspect corrected count rate.
Mathematically, this will manifest itself as a broadening or smearing
of the delta function in \eq{deltacov} by an amount that depends upon
the aspect uncertainties.  The most straightforward way to handle this
effect is to absorb the uncertainties into the PSF itself.  For this
reason, in the following $\F_A(\y, \p',\p)$ will denote the PSF that
includes the effect of the aspect uncertainties.

%}}}

\section{Derivation of the ARF} %{{{
\label{sec:arfdef}

%We shall see how this response may be used to determine the source
%distribution $S(\y,\p)$ from the measured count rate
%$S_D(h,\sigma,t)$. Unfortunately, determining $S(\y,\p)$ the most
%general problem of a spatially extended source is extremely difficult.
%Hence we will start by considering the simplest problem, namely that
%of a single point source.  In subsequent sections, we will deal with
%the more difficult cases.  One should keep in mind that this kind of
%analysis is statistical owing to the fact that the response is a
%probability distribution.

The total number of expected counts with pulse-height $h$ over an
observation interval $\tau$ in some region $\Omega$ can be computed by
integrating $t$ over the observation interval and $\p'$ over the sky
region $\Omega$, i.e.,
\begin{eqnarray}
   C_{\Omega}(h) &=& \int_{\Omega}\D\p'\int_0^{\tau} \d{t} \;
                     S_A(h,\p',t) \\
      &=& \nonumber
      \int\d\y \int_{\Omega} \D\p'\int_0^{\tau} \d{t} \;
          T(\sigma(\p'_t,t),t) D(\sigma(\p'_t,t),h,\y)
            \F_A(\y,\p'_t,\q_t) M(\y,\q_t) S(\y,\p).
\end{eqnarray}
By using \eq{rmfdef} and assuming for the moment that the RMF does not
vary with position, the previous equation can be rewritten as
\begin{equation}
   \label{imageeq}
  C_{\Omega}(h) = \teff \int\d\y\; D_R(h,\y) \int \D\p\; A_{\Omega}(\y,\p)
    S(\y,\p),
\end{equation}
where 
\begin{equation}
   \label{arf}
    A_{\Omega}(\y,\p) = \frac{1}{\teff} \int_{\Omega} \D\p'
        \int_0^{\tau} \d{t}\;
          T(t) Q(\y,\sigma(\p'_t,t)) \F_A(\y,\p'_t,\p_t)M(\y,\p_t)
\end{equation} 
and the effective exposure time is given by
\begin{equation}
   \label{teffdef}
   \teff = \int_0^{\tau} \d{t}\; T(t).
\end{equation}
For simplicity it has been assumed that the good-time interval
function $T(t)$ does not depend upon detector position.

Equation \ref{arf} defines the ARF.  It has the units of area $\times$
counts per photon and depends upon the region $\Omega$, wavelength
$\y$, {\em and} sky position $\p$. However, for the purposes of
point-source analysis, knowledge of the ARF is required only at the
position of the point source, where it may be regarded as depending
only upon wavelength with the understanding that it is valid only for
that source position and region.  But for arbitrary sources, it should
be regarded as an explicit function of $\p$.

% FIXME!!! Spatial information is lost!
Armed with the ARF, \eq{imageeq} is an integral equation that may be
``solved'' to yield the source distribution $S(\y,\p)$ from the
observed aspect-corrected counts $C_{\Omega}(h)$.  Actually, because
of the integration over the sky region $\Omega$, much of the spatial
dependence in $S(\y,\p)$ will be lost and in practice one will have to
assume a known spatial dependence; the examples below illustrate this
point more fully.  One should also realize that the kernel of
\eq{imageeq} is really a probability distribution and that the
observed number of counts will most likely differ from the expected
number of counts predicted by the equation. This implies that
\eq{imageeq} does not really have a unique solution (for a finite
observation time), and any method of ``solving'' should allow for
fluctuations in the number of counts.  One must also take into account
any external background sources as well as any internal background
produced by, say, noise in the detector.  Techniques for treating this
equation are beyond the scope of this paper and may be found elsewhere
\citep{xspecmanual, kahn}.

%The reader
%is referred to \citep{xspecmanual} for information about how the
%popular spectral analysis program {\tt xspec} deals \eq{xspeceq}.

In the next few sections, \eq{imageeq} is considered in the context of
various source distributions.  The problem of the practical
computation of the ARF is taken up in the appendix.

%}}}

\subsection{A Single Point Source} %{{{
A point source located at position $\q$ in the sky with a spectrum
$s(\y)$ may be represented using the source distribution
\begin{equation}
     S(\y,\p) = s(\y)\delta(\p - \q).
\end{equation}
Substituting this equation into \eq{imageeq} yields
\begin{equation}
   \label{xspeceq}
   C_{\Omega}(h) = \teff \int \d\y\; D_R(h,\y) A_{\Omega}(\y,\q) s(\y).
\end{equation}
This equation is essentually the integral equation that the popular
spectral analysis program \verb|xspec| \citep{xspecmanual} is designed
to solve. As noted above, the ARF is required to be computed only at
the position $\q$ of the point source; due to the time-dependence of
the telescope motion, the integration over $t$ in \eq{arf} does
however sample various detector regions and variations in the off-axis
effective area.

%Equation \ref{xspeceq} is an integral equation that relates the
%``unknown'' source flux $s(\y)$ to the ``expected'' number of counts
%in the region $\Omega$ and pulse-height $h$.  A realistic ``solution''
%of \eq{xspeceq} must also take into account any external background
%sources as well as any internal background produced by, say, noise in
%the detector.  Of course, even after the inclusion of these background
%effects, the actual observed number of counts will almost certainly 
%differ from the expected number, and any method of solution will have
%to take such statistical fluctuations into account.

%}}}

\subsection{Multiple Point Sources} %{{{

Multiple point sources may be represented by a source distribution of the
form
\begin{equation}
   S(\y,\p) = \sum_i s_i(\y) \delta(\p - \q_i) ,
\end{equation}
where the position of the $i$th source is given by $\q_i$ and its
spectrum is $s_i(\y)$.  Insertion of this distribution into
\eq{imageeq} produces
\begin{equation}
   C_{\Omega}(h) = \teff \int \d\y\; D_R(h,\y) 
         \bigg[\sum_i A_{\Omega}(\y,\q_i) s_i(\y) \bigg] .
\end{equation}

\begin{figure}
%\epsscale{0.8}
\plotone{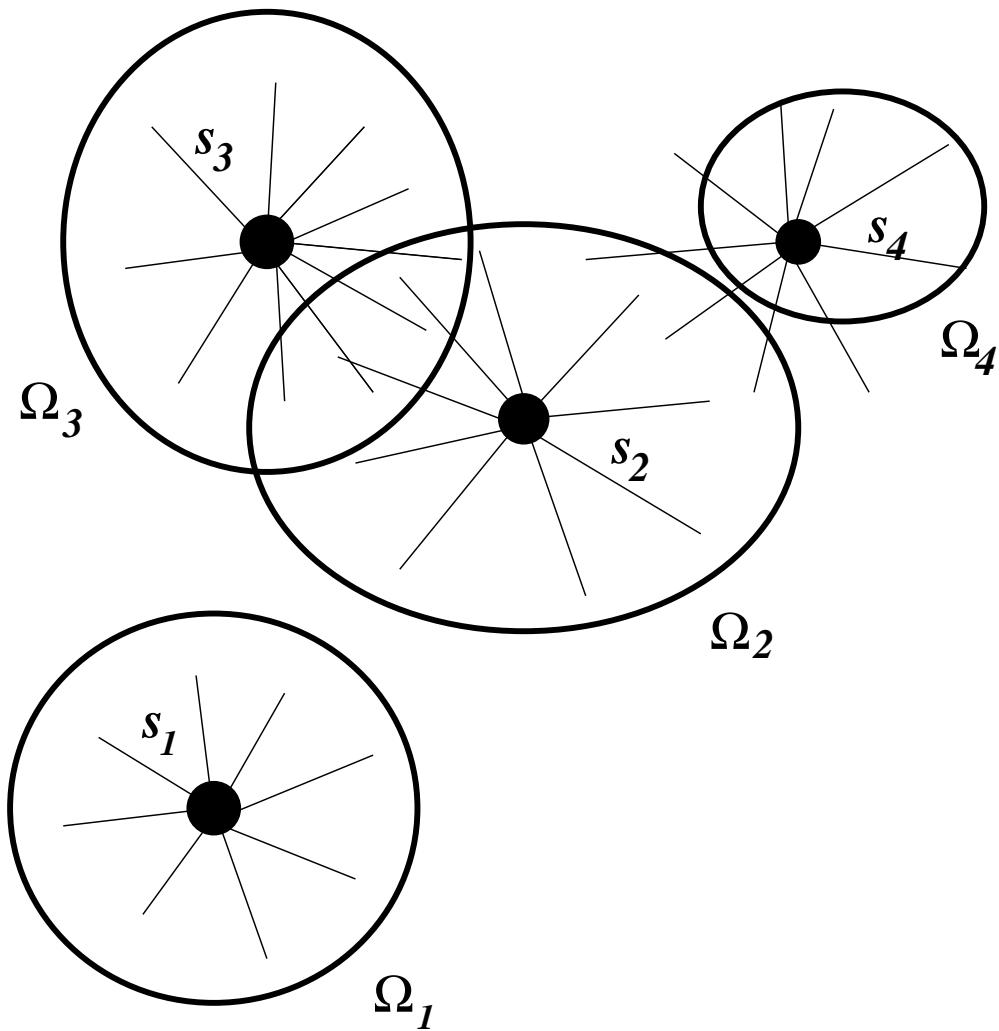}
\caption{%
\label{fig:point}
Figure illustrating the use of multiple regions in \eq{intsys} for a
system of four point sources. In
this figure, the thin radial lines emanating from each of the
point sources represent the PSF (100\% enclosed power) for the
corresponding source, and the elliptical borders outline regions
$\Omega_i$ over which counts are summed.  The PSF from source $s_1$
does not contribute to regions $\Omega_2$, $\Omega_3$, and $\Omega_4$;
hence, it effectively decouples from the other sources. Similarly,
$s_4$ may be treated by itself since none of the other sources
contribute any counts to $\Omega_4$.  However, in dealing with $s_2$,
the contributions from both $s_3$ {\em and} $s_4$ must be taken into
account.}
\end{figure}

If all of the sources have an identical source spectrum such that
$s_i(\y)=s(\y)$, then the resulting integral equation reduces to the
case of a single point source, i.e.,
\begin{equation} 
   C_{\Omega}(h) = \teff \int \d\y\; D_R(h,\y)
         \bigg[\sum_i A_{\Omega}(\y,\q_i)\bigg] s(\y) .
\end{equation}
However, the more interesting case of spectrally distinct sources is
more complicated to solve.  In fact, its solution would require $N$
integral equations since there are $N$ unknown spectral distributions
$s_i(\y)$.  The most straightforward way to obtain the required number
of independent equations would be to use $N$ different regions
$\Omega_i$, not necessarily disjoint, and solve the resulting linear
system of equations
\begin{equation}
   \label{intsys}
   C_{\Omega_i}(h) = \teff \int \d\y\; D_R(h,\y)
         \bigg[\sum_j A_{\Omega_i}(\y,\q_j)s_j(\y)\bigg] .
\end{equation}
These equations are ``coupled'' to the extent that the
$A_{\Omega_i}(\y,\q_j)$ for $i\ne j$ are non-zero, i.e., whether or
not source $j$ has a PSF contribution to region $\Omega_i$ (See figure
\ref{fig:point}). 
Such a system of equations may be handled using
\verb|sherpa| \citep{sherpa}, the Chandra Data System spectral
analysis program.

%}}}

\subsection{Extended Source with Uncorrelated Spatial and Spectral
Distributions} %{{{
\label{sec:extended}
One of the simplest examples of an extended source is one in which the
spatial and spectral distributions are uncorrelated.  That is, the
source distribution $S(\y,\p)$ factors according to
\begin{equation}
    S(\y,\p) = s(\y) \rho(\p),
\end{equation}
where $\rho(\p)$ defines the spatial distribution, assumed to be
properly normalized such that
\begin{equation}
    1 = \int\D\p\;\rho(\p).
\end{equation}
Combining this distribution with \eq{imageeq} yields
\begin{equation}
   C_{\Omega}(h) = \teff \int\d\y\;D_R(h,\y) \bigg[\int\D\p\;
   A_{\Omega}(\y,\p)\rho(\p)\bigg] 
   s(\y) 
\end{equation}
Suppose that the form of $\rho(\p)$ is known.  Then, the ARF could be
combined with the known spatial distribution by defining
\begin{equation}
   A_{\Omega}^{(\rho)}(\y) = \int\D\p\; A_{\Omega}(\y,\p) \rho(\p),
\end{equation}
which leads to the {\tt xspec} style equation
\begin{equation}
   C_{\Omega}(h) = \teff \int\d\y\;D_R(h,\y) A_{\Omega}^{(\rho)}(\y) s(\y)
\end{equation}
for the unknown spectral function $s(\y)$.  Of course, this
methodology cannot be used if the spatial distribution $\rho(\p)$ is not
known.  The more general problem is addressed below in section
\ref{sec:exposuremap} of this paper.

%}}}

\subsection{An ARF in the presence of a spatially varying RMF} %{{{

It is important to note that the ARF, given in \eq{arf}, is useful
only when one can disregard spatial variations in the RMF.
Unfortunately, this may not always be possible.  For example, Chandra
ACIS CCDs have a spatially varying response that must be properly
taken into account.  Provided that one wants to stay within the
confines of the existing ARF+RMF paradigm, the only way to properly
handle such cases is to filter the observed events over a region in
detector coordinates where spatial variations in the RMF may be
neglected.  Mathematically, this procedure may be stated as follows.
Let $\Gamma$ denote the region on the detector where the RMF does not
vary.  Then define a filter $F^{\Gamma}(\sigma)$ on this region by
\begin{equation}
   F^{\Gamma}(\sigma) =
     \left\{ \begin{array}{ll}
        1 & \sigma \in \Gamma, \\
        0 & \mbox{otherwise}.
        \end{array}
     \right.
\end{equation}
Multiplication of \eq{expectrate} by this filter, followed by aspect
correction and summing over the sky region $\Omega$ yields
\begin{equation}
  C_{\Omega}^{\Gamma}(h) = \teff \int\d\y\; D_R^{\Gamma}(h,\y)
     \int \D\p\; A_{\Omega}^{\Gamma}(\y,\p) S(\y,\p),
\end{equation}
where
\begin{equation}
   \label{arfgamma}
    A_{\Omega}^{\Gamma}(\y,\p) = \frac{1}{\teff} \int_{\Omega} \D\p'
        \int_0^{\tau} \d{t}\;
          T(t) F^{\Gamma}(\sigma(\p'_t,t)) Q(\y,\sigma(\p'_t,t)) 
             \F_A(\y,\p'_t,\p_t)M(\y,\p_t).
\end{equation}
In these equations, $C_{\Omega}^{\Gamma}(h)$ is the expected number
of counts in the sky region $\Omega$ that also falls within the detector
region $\Gamma$, $A_{\Omega}^{\Gamma}(\y,\p)$ is the ARF appropriate
for this region of the detector, and $D_R^{\Gamma}(h,\y)$ is the
region-dependent RMF.

%}}}

\section{Definition of the Exposure Map} %{{{
\label{sec:exposuremap}
The ARF presented in the previous section is primarily of use for
spectral analysis over small spatial regions, e.g., the analysis of
point sources.  Much of the spatial information useful for the
treatment of extended sources was lost in the construction of the ARF
by integrating the response over a region $\Omega$ of the sky. A
related product, the exposure map, does not depend upon the
integration over a sky region permitting it to be used for certain
types of extended source analysis.  The goal of this section is to
define an exposure map and show how it may be used with extended
sources.

Start by integrating \eq{aspdatrelat} over an observation time $\tau$
to produce
\begin{eqnarray}
   \label{totaspcounts}
     C(h,\p') &=& \int_0^{\tau} \d{t}\;S_A(h,\p',t) \\
              &=& \nonumber
              \int_0^{\tau} \d{t}
           \int \d\y\; 
              T(\sigma(\p'_t,t),t) D(\sigma(\p'_t,t),h,\y)
              \int \D\p\; \F_A(\y,\p'_t,\p_t) M(\y,\p_t) S(\y,\p).
\end{eqnarray}
Here, $C(h,\p')$ represents the expected total number of
aspect-corrected counts with pulse-height $h$ attributed to the
sky position $\p'$.

In general the PSF is small for rays not too far off the optical axis,
although it can become quite large for far off-axis rays.  Suppose
that the pointing motion amplitude is small enough that the PSF may be
regarded as a scalar under the motion, i.e.,
\begin{equation}
    \label{psfscalar}
    \F_A(\y,\p',\p) = \F_A(\y,\p'_t,\p_t),
\end{equation}
and then consider the integration over $\p$ in \eq{totaspcounts}.  For
off-axis positions where the size of the PSF is small, for 
a {\em fixed} $\p'$, only a narrow range of $\p$ contributes to the
integral.  For the moment, assume that the mirror effective area does
not vary much over this range.
Then the approximation
\begin{equation}
   \int \D\p\; \F_A(\y,\p'_t,\p_t) M(\y,\p_t) S(\y,\p)
       \approx M(\y, \p'_t)\int \D\p\; \F_A(\y,\p',\p) S(\y,\p) 
\end{equation} 
can be used in \eq{totaspcounts} to yield
\begin{equation}
   \label{totaspcounts1}
  C(h,\p') = \int \d\y\; \int_0^{\tau} \d{t}\;
               T(\sigma(\p'_t,t),t) D(\sigma(\p'_t,t),h,\y) M(\y,\p'_t)
                 S_{\F}(\y,\p'),
\end{equation}
where a PSF-smeared source $S_{\F}(\y,\p')$ has been defined by
\begin{equation}
    S_{\F}(\y,\p') = \int \D\p\; \F_A(\y,\p',\p) S(\y,\p) .
\end{equation}
With the introduction of the {\em exposure map} $E(h,\y,\p)$ defined by
\begin{equation}
   \label{emapdef}
   E(h,\y,\p) = \frac{1}{\teff}\int_0^{\tau} \d{t}\;
      T(\sigma(\p_t,t),t) D(\sigma(\p_t,t),h,\y) M(\y,\p_t),
\end{equation} 
\eq{totaspcounts1} may be recast as 
\begin{equation} 
    \label{emapinteq}
   C(h,\p) = \teff \int\d\y\; E(h,\y,\p) S_{\F}(\y,\p).
\end{equation} 
It is important to understand that this is an integral equation
describing the PSF-smeared source and not the true source.  After
``solving'' this equation, one still has the task of removing the
effects of the PSF to determine the true source.  Nevertheless,
\eq{emapinteq} does have one very important feature not shared by the
equations involving the ARF; namely, the spatial distribution of the
expected aspect corrected counts, $C(h,\p)$, is the same as the
PSF-smeared source's spatial distribution.  
%One must also realize that
%\eq{emapinteq} is valid only as long as the size of the PSF is small
%enough that any variation in the effective area over the PSF can be
%neglected.  For far off-axis sources, this equation may not be valid.

%In the following, we shall consider the various uses of the exposure
%map in the context of \eq{emapinteq}.  After that, we shall consider
%the question of the practical computation of \eq{emapdef}.
%\subsubsection{Flux-Correction of Images}

A common use of the exposure map is to remove instrumental
artifacts in images to obtain a better looking image.  This is also
known as ``flux-correcting'' the image.  The method essentially
assumes that the pulse-height resolution of the detector permits the
separation of counts originating from photons from different energy
bands, supplemented by the assumption that the source flux may be
regarded as constant within a band \citep{snowden}.
To express this idea in quantitative terms, consider a range of $\Delta h$ of
pulse-heights centered on some pulse-height $h$ and assume that
the RMF $D_R(h,\y)$ is such that only those photons from the wavelength band
$\Delta\y$ about $\y$ can produce pulse-heights in the specified range.
Now sum \eq{emapinteq} over this range and consider only the photons
from the wavelength band $\Delta\y$ to yield
\begin{equation}
    \sum_{h\in\Delta h} C(h,\p)
       = \teff \int_{\y\in\Delta \y} \d{\y} \;
            \sum_{h\in \Delta h} E(h,\y,\p) S_{\F}(\y,\p),
\end{equation}
which may be written in the more compact form
\begin{equation} 
    C(\Delta h,\p) = \teff \int_{\y\in\Delta \y} \d{\y} \;
            E(\Delta h,\y,\p) S_{\F}(\y,\p),
\end{equation}
with the understanding that the pulse-height range $\Delta h$ is
summed over.

If the bandwidth $\Delta \y$ is such that $E(\Delta h,\y,\p)$ does not vary
much over the band, then it may be removed from the integrand to obtain
\begin{equation}
   \label{emapuse}
       \int_{\y\in\Delta\y} \d\y\; S_{\F}(\y,\p) \approx 
          \frac{1}{\teff} \cdot 
          \frac{C(\Delta h,\p)}%
               {E(\Delta h,\y,\p)}.
\end{equation}
This equation says that the integrated profile of the PSF-smeared
source flux over the wavelength band may be obtained by dividing the
exposure map into the counts image constructed from the appropriate
pulse-height range.  

The resolution of the source spectrum obtained by this technique is
generally poorer than that of the detector's energy resolution because
the wavelength band $\Delta\y$ must be large enough to cover all the
wavelengths that could contribute to the range of pulse-heights
$\Delta h$.  At the same time it must be small enough to ensure that
$E(\Delta h,\y,\p)$ may be treated as a constant in the wavelength
band. It is possible that there may be bands in which these
constraints are mutually exclusive.  For this reason, the use of the
exposure map is limited to situations where spectral resolution is of
secondary importance. For example, spatial resolution is much more
important than spectral resolution when doing source detection.  For
such a situation, one would run the source detection algorithm on an
image obtained by dividing the total counts image over an exposure map
integrated over the bandpass of the telescope.  Another application of
the exposure map would be to use it to get a crude estimate of the
true spectrum, and use that as the first approximation in some more
refined technique.

Before leaving this section, it is important to point out that
\eq{emapinteq} is valid only as long as the size of the PSF is small
enough that any variation in the effective area over the PSF can be
neglected.  If this is not the case, then it is impossible to give a
definition of an exposure map that has the simple relationship between
the observed counts image and the PSF-smeared source as described by
this equation.  By implication it follows that such an exposure map
cannot be used for flux correction via the simple division of
\eq{emapuse}.  However, if the
detector response $D(\sigma,h,\y)$ is uniform, then it is possible to
commute the response with $\F(\y,\p',\p)$ in \eq{totaspcounts} to
produce
\begin{equation}
   C(h,\p') = \teff \int\D\p \int\d{\y} \; 
      \F_A(\y,\p',\p) E(h, \y, \p) S(\y,\p).
\end{equation}
This means that one must first deconvolve the effects of the PSF {\em
before} correcting with the exposure map.  The feasability of this
will depend upon the energy dependence of the PSF where one may have
to perform the deconvolution in specific energy bands.  This
prescription resembles the one advocated by \cite{white} for the
analysis of ASCA data.  Alternatively, \cite{ikebe} has argued that
one start essentually from \eq{totaspcounts} and employ a
``forward-folding'' method to estimate the source distribution
$S(\y,\p)$.

%}}}

\section{Derivation of the Grating ARF} %{{{
\label{sec:garf}
When diffraction gratings are added, the subsystem factorization, 
\eq{imagingrsp}, must be modified to
\begin{equation}
  \label{gratingresponse}
  R(h, \sigma, \y, \p,t) = 
     D(h, \sigma, \y) 
       \int\D\p' \; \delta(\sigma - \sigma_G(\p',t))
        \bigg[\sum_m g_m(\y) \F_m(\y,\p',\p)\bigg] M(\y,\p),
\end{equation}
where $m$ represents the diffraction order and
$\delta(\sigma-\sigma_G(\p',t))$ symbolizes the coordinate transformation
of a diffracted ray at the grating node with direction $\p'$ to the
detector coordinate $\sigma$.  The actual diffraction into the $m$th
order is represented by the term $g_m(\y)\F_m(\y,\p',\p)$, which gives the
probability for a ray with direction $\p$ and wavelength $\y$ to
diffract into the $m$th order with direction $\p'$.  The function
$g_m(\y)$ is the $m$th order grating efficiency, and by definition the
redistribution function satisfies the normalization condition
\begin{equation}
    1 = \int \D\p'\; \F_m(\y,\p',\p).
\end{equation}

\begin{figure}
\epsscale{1.0}
\plotone{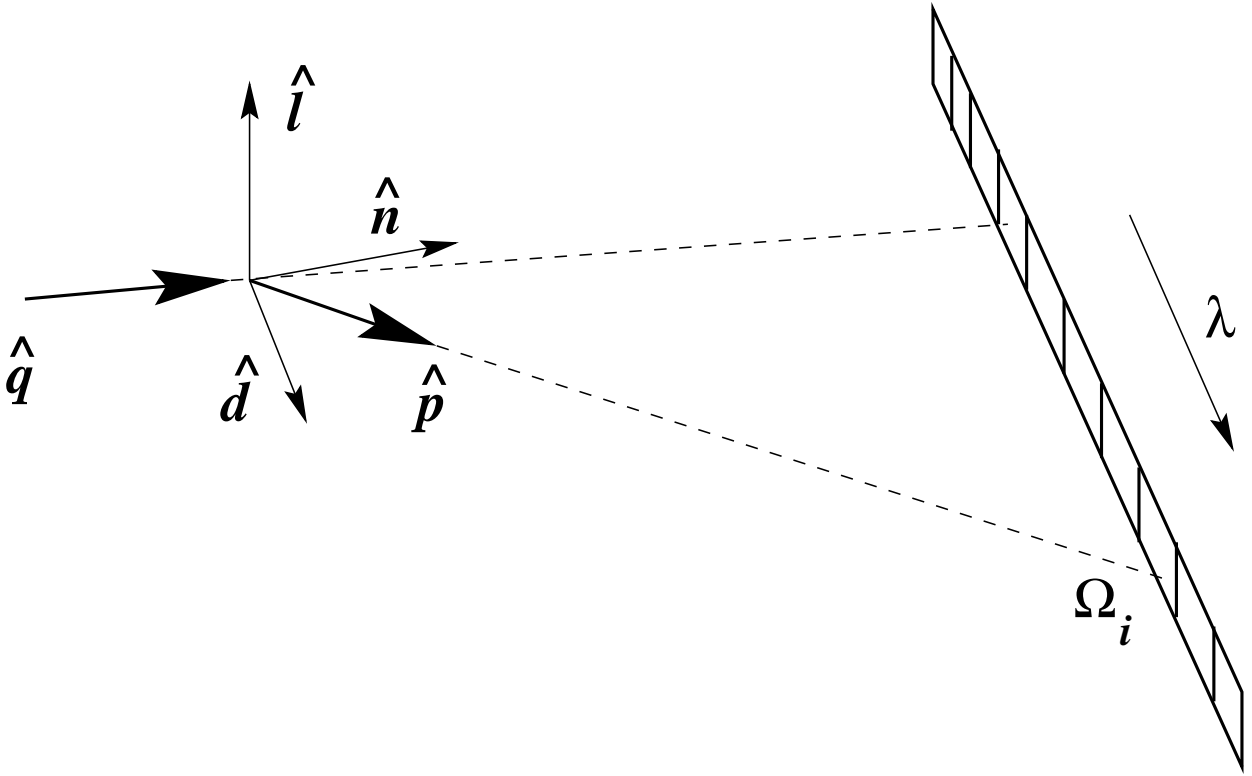}
\caption{%
\label{fig:grating}
Figure illustrating how the grating RMF, $G_{\Omega_i}^{(m)}(\y,\q)$,
represents a redistribution from wavelength $\y$ to regions $\Omega_i$.  In
this figure, the triad of unit vectors $(\vhat{n},\vhat{l},\vhat{d})$
specify an orthonormal coordinate basis centered upon the diffraction grating,
with $\vhat{n}$ normal to the surface of the grating and
$\vhat{l}$ in the direction of the grating bars.  For a perfect
grating, a photon with direction $\q$ and wavelength $\y$ will diffract
into the direction $\p$ in accordance with \eq{gratingeq2} and a
specified diffraction order $m$.  For a realistic grating, the diffraction
process must be described by a probability distribution
$\F_m(\y,\p,\q)$ that is sharply peaked around the set of values that
satisfy \eq{gratingeq2}.  The $m$th order grating RMF,
$G_{\Omega_i}^{(m)}(\y,\q)$ defined by \eq{gratingrmf}, represents the
redistribution probability for an incoming photon with direction $\q$
and wavelength $\y$ to diffract into a region $\Omega_i$.  In other
words, for a specified zeroth order direction $\q$, the grating RMF
may be regarded as a mapping from wavelength $\y$ to region $\Omega_i$
in much the same way as the detector RMF represents a mapping from
wavelength $\y$ to pulse-height $h$. }
\end{figure}

Despite the similarity in form of $\F_m(\y,\p',\p)$ to the imaging PSF
$\F(\y,\p',\p)$, it is important to appreciate one very important
difference between these two functions.  The imaging PSF is sharply
peaked about the set of directions $\p'$ near $\p$.  However,
$\F_m(\y,\p',\p)$ is sharply peaked about a set of directions $\p'$
that vary linearly with wavelength $\y$ according to the diffraction
equation
\begin{equation}
    \label{gratingeq}
    (\p' - \p) \cross \vhat{n} = \frac{m\y}{d} \vhat{l},
\end{equation}
where $d$ is the grating period, the vector $\vhat{n}$ is normal to the
plane of the grating, and $\vhat{l}$ is in the direction of the grating
bars (see figure \ref{fig:grating}).

It follows trivially from \eq{gratingresponse} that the expected
count-rate into $m$th order is given by
\begin{eqnarray}
\label{expectgrate}
       S_D^{(m)}(h, \sigma, t) =
        T(\sigma,t) 
           \int \d\y 
           & & \!\!\!\!\!\!\!\!\!\! 
           \bigg[
              D(h,\sigma,\y) g_m(\y) \\
           & & \!\!\!\!\!\!\!\!\!\! 
         \times \int\D\p'\;
              \delta(\sigma-\sigma_G(\p',t)) 
                 \int\D\p\;\F_m(\y,\p',\p_t) M(\y,\p_t) S(\y,\p) \bigg]
                 \nonumber .
\end{eqnarray}
As in the imaging case, an aspect-corrected count-rate may be defined by
\begin{equation}
   S_A^{(m)}(h,\p,t) = J(\sigma(\p_t,t),\p_t,t) S_D^{(m)}(h, \sigma(\p_t,t), t),
\end{equation}
with the result
\begin{equation} 
     S_A^{(m)}(h,\p', t) =
        T(\sigma_G(\p'_t,t),t) 
           \int \d\y\;
               D(\sigma_G(\p'_t,t),h,\y) g_m(\y) \int\D\p\;
                 \F_m(\y,\p'_t,\p_t) M(\y,\p_t) S(\y,\p).
\end{equation}

This equation may be simplified for the special case of a point source
located at $\q$ with a spectrum $s(\y)$, i.e.,
\begin{equation}
    S(\y,\p') = s(\y)\delta(\p'-\q).
\end{equation}
In addition, assume that the telescope pointing motion amplitudes are
small enough that the grating redistribution function behaves like a
scalar under the motion as in \eq{psfscalar}. Then, the integral over
$\p'$ may be readily performed to yield
\begin{equation}
     S_A^{(m)}(h,\p, t) =
        T(\sigma_G(\p_t,t),t) 
           \int \d\y\;
               D(\sigma_G(\p_t,t),h,\y) g_m(\y)
                 \F_m(\y,\p,\q) M(\y,\q_t) s(\y).
\end{equation}
Integration of this equation over a set of regions $\Omega_i$ and time $\tau$
yields for the total expected number of counts with pulse-height $h$
in the regions
\begin{equation}
   C_{\Omega_i}^{(m)}(h) =
      \int \d\y\;
         \int_{\Omega_i}\D\p \; 
          \bigg[ g_m(\y) 
         \int_0^{\tau} \d{t} \;
            T(\sigma_G(\p_t,t),t)
               D(\sigma_G(\p_t,t),h,\y)
                 M(\y,\q_t) \bigg] \F_m(\y,\p,\q) s(\y).
\end{equation}
For fixed $\y$ and $\q$, the sharp peaked nature of the grating
redistribution function $\F_m(\y,\p,\q)$ implies that only a very
narrow set of directions $\p$ will contribute to the term in square
brackets. Moreover, any telescope pointing motion is expected to
smooth out any non-uniformities in the detector QE appearing in this
term such that one can evaluate it using the value of $\p$ determined
by the grating equation.  In other words, the term in square brackets
can be replaced by a function $A_m(h,\y)$ defined by
\begin{equation}
    \label{gratingarf}
    A_m (h,\y) = g_m(\y) 
         \frac{1}{\teff} \int_0^{\tau} \d{t} \;
            T(\sigma_G(\p_t,t),t)
               D(\sigma_G(\p_t,t),h,\y)
                 M(\y,\q_t),
\end{equation}
where $\p$ satisfies
\begin{equation}
    \label{gratingeq2}
    (\p - \q) \cross \vhat{n} = \frac{m\y}{d} \vhat{l}.
\end{equation}
Hence, the number of counts in the $m$th order with pulse-height $h$
is expected to be
\begin{equation}
  \label{gxspec}
   C_{\Omega_i}^{(m)}(h) =
      \teff \int \d\y\; G_{\Omega_i}^{(m)}(\y,\q) A_m(h,\y) s(\y),
\end{equation}
where
\begin{equation} 
   \label{gratingrmf}
   G_{\Omega_i}^{(m)}(\y,\q) = \int_{\Omega_i}\D\p\;\F_m(\y,\p,\q),
\end{equation}
and $\teff$ is given by \eq{teffdef}.

For reasons that will soon become clear, $A_m(h,\y)$ is called the
grating ARF, and $G_{\Omega_i}^{(m)}(\y,\q)$ is called the grating RMF.
To see this, consider the meaning of \eq{gratingrmf}.  For fixed $\y$,
\eq{gratingrmf} represents a redistribution from
wavelength $\y$ to the region $\Omega_i$, which may be regarded as the
$i$th bin in $\p$-space (see figure \ref{fig:grating}).  This is the
analog of the imaging RMF which describes a redistribution
from $\y$ to a bin in pulse-height space.  With this interpretation,
\eq{gxspec} is formally identical the to equation \eq{xspeceq},
provided that one identifies $A_m(h,\y)$ with the ARF.  Hence, any
techniques that are applicable to \eq{xspeceq} may be readily applied
to the solution of
\eq{gxspec}.

Although $A_m(h,\y)$ depends upon the pulse-height, in practice events
will be filtered upon the pulse-height in order to perform order
separation, provided that the intrinsic energy resolution of the
detector is adequate.  For detectors with poor energy resolution, some
other means of identifying $m$th order events will have to be used.
In any case, $A_m(h,\y)$ will most likely be summed over the range of
pulse-heights appropriate to $m$th order events.  In fact, the
pulse-height range will generally vary with the wavelength such that
the quantity
\begin{equation}
    \label{summedgratingarf}
    A_m(\y) = \sum_{h=h_0(\y)}^{h_1(\y)} A_m(h,\y)
\end{equation}
will actually be what is used in practice.  For this reason, it is
preferable to define the summed quantity $A_m(\y)$ as the grating ARF.

%}}}

\section{Conclusion}

In this paper, explicit expressions for the imaging ARF, grating ARF,
and the exposure map were given in terms of the underlying instrumental
responses that are consistent with the current use of
the objects.  These quantities were obtained from first principles by
relating the expected detector count-rate to an incident photon source
flux via the overall telescope response function suitably
factored into individual instrumental responses.

One of the complications in the derivation of these quantities
concerned the proper treatment of time-varying effects due to
telescope pointing motion, e.g., dither.  At the same time, the
assumed presence of motion about some nominal pointing allowed some
important factorizations to take place that otherwise would have been
suspect in regions containing detector boundaries or bad pixels.  For
this reason, purposely dithering an observation is recommended,
provided, of course, that one can reconstruct the aspect history with
sufficient accuracy.

An added benefit of the first principles approach taken here is that
it allows one to consider problems that cannot readily be handled by
conventional means through the use of an ARF and an RMF. 
For example, as shown
in section \ref{sec:extended}, it ARF may be applied to the
analysis of an extended source provided one knows {\em a priori} that
the source flux distribution factors into a known spatial component
and an unknown spectral component.  It is easy to find sources where
such a factorization is not permissible; the supernova remanent,
Cassiopeia A, is one.  Another problem that does not appear to be
treatable through standard the techniques is the analysis of an
extended source in the presence of a diffraction grating. The grating
ARF defined in section \ref{sec:garf} was derived assuming a point
source distribution.  The basic problem with the analysis of an
extended source is that, unlike a point source, there is no unique
zeroth order position that one could use in the grating equation.  By
judicious filtering in pulse-height space, one may find regions where
there is enough of a point-like behavior to permit the grating ARF to
be used. However, how to handle a generic extended source in the
presence of a diffraction grating is still an open question.  It is
hoped that the mathematical formulation of the extended source problem
as given in sections \ref{sec:arfdef} and \ref{sec:garf} will lead to
better insights into these problems and ultimately to their solution.

This work also highlights some important practical considerations that
should be taken into account in the design of astronomical data
analysis software systems.  For instance, to allow for the possibility
of spatial variation in the underlying detector redistribution
function, the software component responsible for the filtering of
events should allow the user to easily filter {\em simultaneously} on
both sky coordinates and detector coordinates.  In addition, both
filters would need to be passed to the program that generates the ARF.
Finally, spectral fitting programs should be enhanced to facilitate
the analysis of blended sources by handling the coupled integral
equations in \eq{intsys}.

\acknowledgments
I am especially grateful to David Huenemoerder for clarifying a number
of issues during the course of this work.  In addition, I also thank
David Davis, John Houck, Norbert Schulz, and Michael Wise for useful
discussions.  Finally, I am indebted to Dan Dewey for his critical
reading of the text and valuable suggestions regarding the
presentation of some of the material.  This work was supported under
Chandra X-Ray Center contract SV1-61010 from the Smithsonian
Institution.

\appendix

\section{Numerical Considerations}

In this appendix, some ``approximations'' used for the practical
computation of the ARF and grating ARF are discussed.  In fact, these
approximations are actually employed by the Chandra exposure map code
suite for the generations of exposure maps and ARFs.  Since the code
is freely available\footnote{%
See {\tt http://chandra.harvard.edu} for more information. }, the
actual implementation details will not be discussed here.  The reader
should also note that some of these approximations may only be valid
for the Chandra telescope, which dithers, and for other missions one
may have to resort to the full definitions given in the main body of
the text.

\subsection{Performing the Time Integrations via an Aspect Histogram}

The integrals over the observation time appearing in the equations for
the ARF and the grating ARF can be quite computationally expensive,
especially for long observation times.  The
general form of these integrals is given by
\begin{equation}
   \label{timeinteg}
   I = \int_0^{\tau} \d{t}\; T(t) F(\vect{\alpha}(t))
\end{equation}
where $\vect{\alpha}(t)$ is an $N$ dimensional time-dependent vector
that characterizes the dither of the telescope and the relative motion
of its subsystems.  For example, $N$ is $3$ if there is no internal
movement, and the dither is characterized by the roll, pitch, and yaw
of the telescope.

By multiplying the preceeding equation by the identity
\begin{equation}
   1 = \int\d\vect{\alpha} \; \delta(\vect{\alpha} - \vect{\alpha}(t))
\end{equation} 
it trivially follows that
\begin{equation} 
   \label{asphistuse}
   I = \int\d\vect{\alpha} {\cal H}(\vect{\alpha}) F(\vect{\alpha}),
\end{equation} 
where
\begin{equation} 
   {\cal H}(\vect{\alpha}) = \int_0^{\tau} \d{t}\;T(t) \delta(\vect{\alpha} - \vect{\alpha}(t)).
\end{equation}
The quantity ${\cal H}(\vect{\alpha})\d\vect{\alpha}$ has a very simple
interpretation. It represents the total amount of time, weighted by
$T(t)$, that the point spent in the volume element
$\Delta\vect{\alpha}$ at $\vect{\alpha}$.  Now, if the telescope dithers
around some nominal pointing, and if the time-dependent internal
motions due to, e.g., thermal expansion are small, then the point
$\vect{\alpha}(t)$ will be confined to some small volume in the $N$
dimensional space.  This means that ${\cal H}(\vect{\alpha})$ will be
non-zero only in that small volume and zero everywhere else. So, to
compute the time integration over long observation times for the case
of small dither amplitudes, it is often more efficient to compute the value of
${\cal H}(\vect{\alpha})$ and use it to evaluate \eq{asphistuse}.  In
practice, the portion of the $N$ dimensional space where ${\cal
H}(\vect{\alpha})$ is non-zero is sub-divided into small volume
elements $\Delta\vect{\alpha}$. Then the discretized quantity ${\cal
H}_{\vect{\alpha};\Delta\vect{\alpha}}={\cal
H}(\vect{\alpha})\Delta\vect{\alpha}$ is computed and used in a
discretized version of \eq{asphistuse}, i.e.,
\begin{equation}
   \label{asphistapprox}
   I \approx \sum_{\vect{\alpha}} {\cal H}_{\vect{\alpha},\Delta\vect{\alpha}} F(\vect{\alpha})
\end{equation}
For reasons that should be apparent, ${\cal
H}_{\vect{\alpha};\Delta\vect{\alpha}}$ is called the {\em aspect
histogram}.

The major advantage of this approach for the case of small dither
amplitudes is that there are likely to be many fewer terms to sum in
\eq{asphistapprox} than if a straightforward discretization were used
to perform the time integration in \eq{timeinteg}.  Moreover, there
are efficient algorithms based upon $2^N$-ary trees for computing the
aspect histogram.  For example, the code for both the Chandra and
ROSAT missions use an octtree for $N=3$. (For Chandra, the value of
$N$ used is 3 rather than 6 through the use of ``effective''
offsets.)

\subsection{Computation of the Imaging ARF} %{{{

The ARF is a complicated function requiring complete knowledge of the
detector QE, mirror effective area, aspect solution, and the point
spread function.  To compute it directly from
\eq{arf} or from \eq{arfgamma} in the case of a spatially varying RMF,
one would need to carry out an integration over time as well as a 2-d
integration over the sky region, and do this for every point in the
sky.  Clearly, this is not practical and in view of the fact that
there will be uncertainties in the instrumental responses at this
level of detail, such a calculation is unwarranted.  Instead, one can
make several simplifying approximations that permit the ARF to be computed in an
economic manner.

As written, \eq{arf} is valid for any motion of the spacecraft,
including slew.  However, here it shall be assumed that one is
dithering about some mean pointing and that the scale of the dither
is small enough that any variations in the PSF and the mirror
effective area on this scale can be neglected.  Therefore,
\eq{arf} will be approximated by
\begin{equation}
    A_{\Omega}(\y,\p) \approx \frac{1}{\teff} \int_{\Omega} \D\p'
       \F_A(\y,\bra\p'\ket,\bra\p\ket)M(\y,\bra\p\ket) 
       \int_0^{\tau} \d{t}\;
          T(t) Q(\y,\sigma(\p'_t,t)),
\end{equation}
where $\bra\p\ket$ represents the time-average of $\p_t$, i.e.,
\begin{equation}
   \bra\p\ket = \frac{1}{\teff}\int_0^{\tau} \d{t}\; T(t)\R(t)\cdot\p .
\end{equation}
Similarly, define
\begin{equation}
    \label{braQket}
     \bra Q(\y,\p') \ket = \frac{1}{\teff}
        \int_0^{\tau} \d{t}\; T(t) Q(\y,\sigma(\p'_t,t))
\end{equation} 
to be the time-averaged value of the QE.  Then one can write
\begin{equation}
    A_{\Omega}(\y,\p) \approx M(\y,\bra\p\ket) \int_{\Omega} \D\p'
       \F_A(\y,\bra\p'\ket,\bra\p\ket) \bra Q(\y,\p') \ket .
\end{equation}
The time-averaging over the dither motion has the effect of smoothing
out any large variations in the QE over the region.  In fact, this is
the primary purpose of the dither.  Now since $\bra Q(\y,\p')\ket$ can
be assumed to vary slowly over the region, and since $\F_A(\y,\p',\p)$
is expected to rapidly go to zero as $\p'$ moves away from $\p$, 
$\bra Q(\y,\p')\ket$ can be replaced by its average over the region and
removed from the integrand.  This leads to the result
\begin{equation}
     A_{\Omega}(\y,\p) \approx f_{\Omega}(\y,\p) M(\y,\bra\p\ket)
     \bra Q(\y)\ket_{\Omega},
\end{equation} 
where
\begin{equation}
     \bra Q(\y)\ket_{\Omega} = \frac{1}{\Omega} \int_{\Omega} \D\p'\;
         \bra Q(\y,\p') \ket
\end{equation}
is the average of $\bra Q(\y,\p)\ket$ over the region and the PSF
{\em fraction} in the region is given by
\begin{eqnarray}
    f_{\Omega} (\y,\p) 
       & = & \int_{\Omega} \D\p'\; \F_A(\y,\bra\p'\ket,\bra\p\ket) \\ \nonumber
       & \approx & \int_{\Omega} \D\p'\; \F_A(\y,\p',\p) .
\end{eqnarray} 

%}}}

\subsection{Computation of the Grating ARF} %{{{

The grating ARF is defined by equation \eq{summedgratingarf}, rewritten
here as
\begin{equation}
  A_m(\y) = \bigg[\sum_{h=h_0(\y)}^{h_1(\y)} D_R(h,\y)\bigg]
        \times g_m(\y) 
         \frac{1}{\teff} \int_0^{\tau} \d{t} \;
            T(t) Q(\sigma_G(\p_t,t),t) M(\y,\q_t),
\end{equation}
where \eq{rmfdef} has been used with the assumption that the RMF does
not vary spatially.  For a spatially varying RMF an additional
spatial filter would need to be applied as was done for the imaging ARF
to derive \eq{arfgamma}.  In the above equation, $\p$ depends upon the
source position and the wavelength according to
\begin{equation}
    (\p - \q) \cross \vhat{n} = \frac{m\y}{d} \vhat{l}.
\end{equation}
If the amplitude of the dither is small on the scale of the variations in
the mirror effective area, then $M(\y,\q)$ may be replaced by
$M(\y,\bra\q\ket)$ and removed from the integrand.  Hence, the grating
ARF may be approximated by
\begin{equation} 
  A_m(\y) \approx g_m(\y) \bra Q(\y,\p) \ket M(\y,\bra\q\ket) 
       \sum_{h=h_0(\y)}^{h_1(\y)} D_R(h,\y),
\end{equation}
where
\begin{equation} 
   \bra Q(\y,\p)\ket = \frac{1}{\teff} \int_0^{\tau} \d{t} \;
            T(t) Q(\sigma_G(\p_t,t),t).
\end{equation} 

%}}}

\end{document}